\documentstyle[12pt]{article}
\begin{document}

\vspace{15mm}
\begin{center}
{\Large {\bf Relativistic effects in radiative transitions of
charmonia}}
\vspace{1cm}

{\large I.G.Aznauryan} \\
\vspace{1cm}
{\em Yerevan Physics Institute}\\
{(Alikhanian Brothers St. 2, Yerevan, 375036 Armenia)}\\
{(e-mail: aznaurian@vx1.yerphi.am)}
\end{center}

\vspace{5mm}
\renewcommand{\thefootnote}{\arabic{footnote}}
\setcounter{page}{1}

\indent
In the framework of a relativistic quark model
constructed in an infinite momentum frame, corrections
of the order of $v^{2}/c^{2}$ are obtained in the formulas
which express amplitudes of the transitions
$\psi(\psi')\rightarrow\eta_{_{c}}(\eta'_{_{c}})\gamma$,
$\chi_{_{J}}\rightarrow\psi\gamma$,
$\psi'\rightarrow\chi_{_{J}}\gamma$,
$h_{_{c}}\rightarrow\eta_{_{c}}\gamma$,
$\eta'_{_{c}}\rightarrow h_{_{c}}\gamma$ through overlap integrals
$\int\phi^{*}_{f}(q)\phi_{i}(q)d{\bf q}$,
$\int\phi^{*}_{f}(r)r\phi_{i}(r)d{\bf r}$.
These corrections lead to a considerable
suppression of $\Gamma(\psi\rightarrow\eta_{_{c}}\gamma)$,
however, they are
insufficient to remove the existing disagreement
of quark model predictions with experiment.
Taking into account the relativistic
corrections, the ratios of overlap integrals
for
$\chi_{_{0}}\rightarrow\psi\gamma,~\chi_{_{2}}\rightarrow\psi\gamma$
and
$\psi'\rightarrow\chi_{_{0}}\gamma,~\psi'\rightarrow\chi_{_{2}}\gamma$
transitions are extracted from experiment; the obtained
results are compared with quark model predictions.

\newpage

\section{Introduction}

In addition to the mass spectrum of $q\bar {q}$ - states, which is a good
laboratory for understanding of interquark forces,
the quarkonium single-photon
transitions also provide sensitive tests for potential models. Recently
a considerable progress has been achieved in the experimental
investigation of the widths
of charmonium radiative transitions due to the more
precise measurement of the $\psi$ width
[1] and due to the first measurements of
the $\chi_{_{c1}}$ and $\chi_{_{c2}}$
widths [2] as well. The further progress
in the investigation of the charmonium
radiative transitions will be available at
tau-charm factories where
along with more precise measurements of
the $\chi_{_{J}}\rightarrow\psi\gamma$,
$\psi'\rightarrow\chi_{_{J}}\gamma$,
$\psi\rightarrow\eta_{_{c}}\gamma$
decay widths, the measurements of the radiative
transitions of the missing $\eta'_{_{c}}$
and recently discovered $^{1}P_{1}(h_{_{c}})$ [3] states
($\psi'\rightarrow\eta'_{_{c}}\gamma$,
$h_{_{c}}\rightarrow\eta_{_{c}}\gamma$,
$\eta'_{_{c}}\rightarrow h_{_{c}}\gamma$) can be accessed.

The present work is devoted to studying the
relativistic effects in the
above-listed charmonium radiative transitions,
namely, the corrections
of the order of $v^{2}/c^{2}$ will be found in
the formulas which express
amplitudes of these transitions through
the wave-function overlap integrals:
\begin{equation}
I_{_{fi}}=
\int\phi^{*}_{f}(q)\phi_{i}(q)
d{\bf q},~~r_{_{fi}}=
\int\phi^{*}_{f}(r)r\phi_{i}(r)
d{\bf r}.
\end{equation}

The reasons, why such consideration is important,
are following.
(a) The relativistic effects in wave functions
of charmonia
are investigated in many papers (see, e.g., refs.[4-6])
within a two-component model for interquark interactions:
short distance
one-gluon exchange and long-range confinement;
however, to calculate the amplitudes
of radiative transitions in these papers
formulas of nonrelativistic
quantum mechanics are used.
(b) In the quark model predictions
for the overlap integrals (1)
there are results, which depend weakly
on the details in choosing of the potential
and on the method of taking into account
the relativistic corrections in the wave functions.
These are ratios of the overlap integrals
$r_{fi}$ for different states $\chi_{J}$,
which are close to the unity
due to the small spin-orbit splitting of
$\chi_{_{0}}$, $\chi_{_{1}}$, $\chi_{_{2}}$.
Moreover, the analysis of the results
of ref.[4,6] shows that the deviations
of these ratios from the unity are hopefully predicted 
in the quark model. These are the overlap integrals $I_{fi}$
for the transitions
$\psi\rightarrow\eta_{_{c}}\gamma$,
$\psi'\rightarrow\eta'_{_{c}}\gamma$,
which are equal to the unity in
nonrelativistic approximation;
the relativistic corrections do not considerably
affect these integrals due
to the small spin-spin
splitting generated by the
one-gluon exchange in the $L=0$ charmonium systems.
The ratios of the overlap integrals $r_{fi}$
for the transitions
$h_{_{c}}\rightarrow\eta_{_{c}}\gamma$ and
$\chi_{_{J}}\rightarrow\psi\gamma$,
$\eta'_{_{c}}\rightarrow h_{_{c}}\gamma$ and
$\psi'\rightarrow\chi_{_{J}}\gamma$
are close to the unity too,
since the absence of significant long-range spin-spin forces,
as is seen in the near degeneracy of $h_{_{c}}$ and
the $\chi_{_{J}}$ multiplet
c.o.g., leads to near equality of the wave functions for
$h_{_{c}}$ and $\chi_{_{J}}$.
Therefore, for a comparison of these quantities
with experiment it is especially important
to take into account correctly relativistic effects
in the amplitudes.

The relativistic corrections in the amplitudes
of charmonium radiative
transitions, we will investigate within
a relativistic quark model constructed in the light-front
dynamics in refs.[7,8] and formulated later
in an infinite-momentum frame (IMF)
in refs.[9,10]. It turned out that in the considered
approximation (when
only corrections of the order of $v^{2}/c^{2}$
have been taken into account)
the amplitudes of the transitions under interest
are determined by the form factors
which can be found using only linear in photon transverse
momentum terms
of the longitudinal components of electromagnetic transition current.
In ref.[8] it was shown, that the results
obtained using these terms are
consistent with relativistic invariance.

\section{Basic formulas and results}

\indent

The relativistic quark model [9,10] is constructed for the hadron
radiative transitions  $A(P) \rightarrow B(P')+\gamma^*(K)$
(in parentheses the momenta of particles are given)
in the IMF, where the initial hadron moves along
the z axis with momentum  $P \rightarrow \infty$,
and the photon momentum components are ${\bf K}_\bot$,
$K_z=-(m_A^2-m_B^2-{\bf K}_\bot^2)/4P$,
$ K_0=-K_z$,  $K^2=-{\bf K}_\bot^2$.
In this frame (see also ref.[11]), the space-time picture
of the process for the longitudinal
components of the electromagnetic current
taken between hadron states with the helicities:
$\lambda=\pm S_{A},~~\lambda'=\pm S_{B}$,
is the same as in the
nonrelativistic quantum mechanics. The
corresponding matrix elements
have the following form [9,10]:
\begin{equation}
\label{AB}
\frac{1}{2P}\langle P',\lambda^\prime=\pm S_{B}
\mid J_{0,z}\mid P,\lambda=\pm S_{A}\rangle\mid_{P\rightarrow\infty}=
\frac{4}{3}e\int d\Gamma\Phi_{A}(M_{0}^2)\Phi_{B}(M_0'^2)T,
\end{equation}
\begin{equation}
 T=Sp\{U^+(x_{\bar{c}},{\bf q'}_{\bar{c}_{\bot}})
(\Gamma_B^{\lambda^{'}})^{+}
U(x_{c},{\bf q'}_{{c}_{\bot}})U^{+}(x_{c},{\bf q}_{{c}_\bot})
\Gamma_{A}^{\lambda}U(x_{\bar{c}},{\bf q}_{\bar{c}_{\bot}})\}.
\end{equation}

\indent
Here we have assumed, that the hadrons A,B are bound states
$c\bar{c}$,
the quark momenta in the initial and final hadrons
are parametrized in the IMF by:
\begin{eqnarray}
&&{\bf q}_{i}=x_{i}{\bf P}+{\bf q}_{i_{\bot}}, ~~{\bf q'}_{i}=
x_{i}{\bf P'}+{\bf q'}_{i_{\bot}},~i=c,\bar{c},\nonumber \\
&&x_{c}=1-x, ~~x_{\bar{c}}=x, ~~{\bf q}_{c\bot}=-{\bf q}_{\bar{c}\bot}=
-{\bf q}_{\bot},\\
&&{\bf q'}_{c\bot}=-{\bf q'}_{\bar{c}\bot}=
-{\bf q'}_{\bot},~~{\bf q'}_{\bot}=
{\bf q}_{\bot}+x{\bf K}_{\bot}, \nonumber
\end{eqnarray}
$M_{0}$ and $M'_{0}$ are invariant masses of
the systems of initial and
final quarks:
\begin{equation}
M_{0}^{2}=\frac{m^{2}+{\bf q}_{\bot}^{~2}}{x(1-x)},~~{M'_{0}}^{2}=
\frac{m^{2}+{\bf q'}_{\bot}^{~2}}{x(1-x)},
\end{equation}
$m$ is the mass of the $c$ quark. The introduced variables are related
to the 4-momentum of the initial quarks in their CMS
 $q({\bf q}_{\bot},q_{z},\varepsilon)$ by:
\begin{equation}
q_{z}+\varepsilon=M_{0}x,~\varepsilon^{2}={\bf q}^{2}+m^{2},
~M_{0}=2\varepsilon.
\end{equation}
In terms of variables $x$,~${\bf q}_{\bot}$~ and~ ${\bf q}$ ~
the phase space volume $d\Gamma$ has the form:
\begin{equation}
 (2\pi)^{3}d\Gamma=dx~d{\bf q}_{\bot}/2x(1-x)=2d{\bf q}/M_{0}.
\end{equation}
In eqs. (2,3) we have taken into account, in accordance with the
results of refs. [9,10], that the vertex functions
of the hadron transitions
to quarks in the IMF are related to the wave
functions of quarks in their
CMS by spin rotation given by the Melosh matrix:
\begin{equation}
 U(x_{i}, {\bf q}_{i_{\bot}})=
\frac{m+M_{0}x_{i}+i\varepsilon_{lm}\sigma_{l}q_{im}}
 {[(m+M_{0}x_{i})^{2}+{\bf q}_{\bot}^{~2}]^{1/2}};
\end{equation}
$\Gamma$ are the spin-orbit parts of these wave functions
in the CMS of quarks:
\begin{eqnarray}
&&\Gamma_{\eta_{_{c}}}=\frac{i}{\sqrt{2}},
~~\Gamma_{\psi}^{\lambda}=
{\bf \sigma}{\bf e}_{\psi}^{~(\lambda)}/\sqrt{2},~~
~~\Gamma_{h_{_{c}}}=\sqrt{\frac{3}{2}}{\bf n}{\bf e}_
{h_{_{c}}}^{~(\lambda)},\\
&&\Gamma_{\chi_{_{0}}}={\bf \sigma}{\bf n}/\sqrt{2},~~
\Gamma_{\chi_{_{1}}}^{\lambda}=\frac{\sqrt{3}}{2}
[{\bf \sigma}{\bf n}]
{\bf e}_{\chi_{_{1}}}^{~(\lambda)},
~~\Gamma_{\chi_{_{2}}}^{\lambda}=
\sqrt{\frac{3}{2}}(e_{\chi_{_{2}}}^{(\lambda)})_{lm}n_{l}\sigma_{m},
\end{eqnarray}
 (${\bf n}={\bf q}/q$).
$\Phi(M_{0}^{2})$ are radial parts of wave functions
which in the present work we take in the
form corresponding to the oscillator potential:
\begin{eqnarray}
&&\Phi_{\psi(\eta_{_{c}})}(M_{0}^{2})/\sqrt{M_{0}}
\sim exp(-M_{0}^{2}/4\alpha^{2})\sim exp(-q^{2}/
\alpha^{2}),\nonumber\\
&&\Phi_{\chi(h_{_{c}})}(M_{0}^{2})/\sqrt{M_{0}}
\sim q~exp(-M_{0}^{2}/4\alpha^{2}),\\
&&\Phi_{\psi'(\eta'_{_{c}})}(M_{0}^{2})/\sqrt{M_{0}}
\sim (q^{2}-\frac{3}{4}\alpha^{2})exp(-M_{0}^{2}/4\alpha^{2}).\nonumber
\end{eqnarray}
This permits us to
make calculations analytically and obtain the contribution
of relativistic corrections into transition
amplitudes in an explicit form.

The widths of the considered decays are
related to the amplitudes corresponding
to the M1 and E1 transitions via:
\begin{equation}
\Gamma(\psi(\psi')\rightarrow\eta_{_{c}}(\eta'_{_{c}})\gamma)=
\alpha~\frac{16}{27}~\frac{\omega^{3}}{m^{2}}|M1|^{2},
\end{equation}
\begin{equation}
\Gamma(\chi_{_{J}}\rightarrow\psi\gamma)=
\alpha~\frac{16}{81}~\omega^{3}|E1|^{2},
\end{equation}
\begin{equation}
\Gamma(\psi'\rightarrow\chi_{_{J}}\gamma)=
\alpha~\frac{16}{81}~\frac{2J+1}{3}
~\omega^{3}|E1|^{2},
\end{equation}
\begin{equation}
\Gamma(h_{_{c}}\rightarrow\eta_{_{c}}\gamma)=
\alpha~\frac{16}{81}~\omega^{3}|E1|^{2},
\end{equation}
\begin{equation}
\Gamma(\eta'_{_{c}}\rightarrow h_{_{c}}\gamma)=
\alpha~\frac{16}{27}~\omega^{3}|E1|^{2}.
\end{equation}

In the nonrelativistic approximation
these amplitudes are equal directly to
the wave-function overlap integrals:
$M1(nonrel)=I_{_{fi}},~~
E1(nonrel)=r_{_{fi}}$. Let us note, that the functions
$\phi(r)$ and
$\phi(q)\sim
\Phi(M_{0}^{2})/\sqrt{M_{0}}$
are radial parts of wave functions
in the momentum and coordinate representations,
which are normalized by:
\begin{equation}
\int|\phi(r)|^{2}d{\bf r} =
 \int|\phi(q)|^{2}d{\bf q} =
 \int|\Phi(M_{0}^{2})|^{2}d\Gamma = 1.
\end{equation}

In order to take into account the relativistic effects, one should
find relativistic corrections in formulas expressing
the amplitudes M1 and E1 through the overlap integrals
$I_{_{fi}}$ and $r_{_{fi}}$. In addition, the contributions
corresponding to M2 transition in the
$\chi_{_{1}}\rightarrow\psi\gamma$, $\psi'\rightarrow\chi_{_
{1}}\gamma$ decays and to M2 and E3 transitions in the
$\chi_{_{2}}\rightarrow\psi\gamma$, $\psi'\rightarrow\chi_{_
{2}}\gamma$ decays, should be taken into account. These
contributions are suppressed as compared to the contribution
of E1 as $\omega^{2}/m^{2}$, where $\omega=(m_{i}^{2}-m_{f}^{2})/2m_{i}$.
For the oscillator potential, this is the value of the order of
$(\alpha/m)^{4}$. Therefore, in the considered approximation
(when we take into account only corrections of the order of
$v^{2}/c^{2}\sim\alpha^{2}/m^{2})$ these contributions should
be neglected. In this approximation amplitudes M1 and E1
are equal to the form factors corresponding to the following
relativistic - covariant expressions for the matrix elements
of the electromagnetic current:
\begin{eqnarray}
\psi\rightarrow\eta_{_{c}}\gamma:&\langle P'|J_{\mu}|P,
~\lambda\rangle=
~e~\frac{4}{3}M1~\varepsilon_{\mu\nu\sigma\rho}K^
{^{\nu}}P^{\sigma}~e_{(\lambda)}^{\rho},\\
h_{_{c}}\rightarrow\eta_{_{c}}\gamma:&\langle
P'|J_{\mu}|P,~\lambda \rangle=
~e~\frac{4}{3\sqrt{3}}E1~[e_{\mu}^{(\lambda)}(PK)~-
~P_{\mu}(Ke^{(\lambda)})],\\
\chi_{_{0}}\rightarrow\psi\gamma:&\langle P',
~\lambda'|J_{\mu}|P \rangle=
~e~\frac{4}{9}E1~[e_{\mu}^{(\lambda')}(PK)~-
~P_{\mu}(Ke^{(\lambda')})],\\
 \chi_{_{1}}\rightarrow\psi\gamma:&\langle P',
~\lambda'|J_{\mu}|P,~\lambda\rangle=
~e~\sqrt{\frac{32}{27}}~
\frac{m_{\psi}m_{\chi}}
{m_{\psi}+m_{\chi}}E1
~\varepsilon_{\mu\nu\sigma\rho}K^{^{\nu}}e_{(\lambda')}
^{\sigma}~e_{(\lambda)}^{\rho},\\
\chi_{_{2}}\rightarrow\psi\gamma:&\langle P',
~\lambda'|J_{\mu}|P,~\lambda\rangle=
~e~\frac{4}{3\sqrt{3}}E1~[~e_{\mu\nu}^
{(\lambda)}(KP)-P_{\mu}e_{\nu\alpha}^
{(\lambda)}K^{\alpha}]~e_{(\lambda')}^{\nu}.
\end{eqnarray}

It can be easily seen that all these form factors,
except that for the $\chi_{1}\rightarrow \psi \gamma$
transition, can be found using the relations (2,3)
from the following matrix elements:
\begin{equation}
\frac{1}{2P}\langle\lambda_{\eta_{c}}=0|J_{0}|
\lambda_{\psi}=1\rangle
=ie~\frac{\sqrt{2}}{3}(M1)K_{x},
\end{equation}
\begin{equation}
\frac{1}{2P}\langle\lambda_{\eta_{c}}=
0|J_{0}|\lambda_{h_{c}}=1\rangle
=e~\sqrt{\frac{2}{3}}~\frac{E1}{3}K_{x},
\end{equation}
\begin{equation}
\frac{1}{2P}\langle\lambda_{\psi}=1|J_{0}|
\lambda_{\chi_{_{0}}}=0\rangle=
e~\frac{\sqrt{2}}{9}(E1)K_{x},
\end{equation}
\begin{equation}
\frac{1}{2P}\langle\lambda_{\psi}=1|J_{0}|
\lambda_{\chi_{_{2}}}=2\rangle
=e~\sqrt{\frac{2}{3}}~\frac{E1}{3}K_{x}.
\end{equation}

The amplitude E1
for the $\chi_{1}\rightarrow \psi \gamma$
transition can not be found from relations (2,3)
in the leading over P order;
therefore, in the considered approach
we can not obtain reliable results for this transition.
From the relations (23-26)
it is seen, that all $M1$ and $E1$ amplitudes
for remaining transitions
are determined by the linear over $K_{_{x}}$ terms of the matrix
elements (2). To find these terms, one should keep in
$\Phi_{_{B}}(M_{0}^{\prime~2})$
only the terms of the order of $(K_{_{x}})^{0}$ and $K_{_{x}}$:
\begin{equation}
\Phi_{_{B}}(M_{0}^{\prime~2})=\Phi_{_{B}}(M_{0}^{2})\left[1-
\frac{q_{x}K_{x}}{(1-q_{z}/m)\alpha^{2}}
+\frac{q_{x}K_{x}}{4m^{2}}\right],
\end{equation}
and write the quantity $T$ in the form:
$$
T=Sp\{I_{\bar{c}}\left(\Gamma_{_{B}}^{\lambda'}\right)^{+}
I_{c}\Gamma_{_{A}}^{\lambda}\},
$$
where
\begin{equation}
I_{c}\equiv U(x_{c}, {\bf q'}_{c_{\bot}})U^{+}(x_{c},
{\bf q_{c_{\bot}}})=
1+i\frac{K_{x}}{4m}\left[\sigma_{_{2}}\left(1+\frac{q_{z}}{2m}-
\frac{2q^{2}}{3m^{2}}\right) +
\sigma_{_{3}}\frac{q_{y}}{2m}\right],
\end{equation}
\begin{equation}
I_{\bar{c}}\equiv U(x_{\bar{c}}, {\bf q}_
{\bar{c}_{\bot}})U^{+}(x_{\bar{c}},
{\bf q'_{\bar{c}_{\bot}}})=
1+i\frac{K_{x}}{4m}\left[\sigma_{_{2}}\left(1+\frac{3q_{z}}{2m}-
\frac{q^{2}}{3m^{2}}\right) +\sigma_{_{3}}\frac{q_{y}}{2m}\right].
\end{equation}
\indent
Further, using the formulas (2,3,23-29), it is easy to obtain:
\begin{eqnarray}
&&M1(\psi\rightarrow\eta_{c}\gamma)=I_{\eta_{c}\psi}
\left(1-\frac{3\alpha^{2}}{8m^{2}}\right),\\
&&M1(\psi'\rightarrow\eta'_{c}\gamma)=I_{\eta'_{c}\psi'}
\left(1-\frac{33\alpha^{2}}{56m^{2}}\right),\\
&&E1(h_{_{c}}\rightarrow\eta_{c}\gamma)=
r_{\eta_{c}h_{_{c}}},~~E1(\chi_{_{0}}\rightarrow\psi\gamma)=
r_{\psi\chi_{_{0}}},\\
&&E1(\chi_{_{2}}\rightarrow\psi\gamma)=r_{\psi\chi_{_{2}}}
\left(1+\frac{\alpha^{2}}{4m^{2}}\right),\\
&&E1(\eta'_{c}\rightarrow~h_{_{c}}\gamma)=
r_{h_{_{c}}\eta'_{c}}\left(1-\frac{7}{12}
~\frac{\alpha^{2}}{m^{2}}\right),\\
&&E1(\psi'\rightarrow\chi_{_{0}}\gamma)=
r_{\chi_{_{0}}\psi'}\left(1-\frac{7}{12}
~\frac{\alpha^{2}}{m^{2}}\right),\\
&&E1(\psi'\rightarrow\chi_{_{2}}\gamma)=r_{\chi_{_{2}}\psi'}
\left(1-\frac{5}{6}~\frac{\alpha^{2}}{m^{2}}\right).
\end{eqnarray}
\section{Comparison with experiment and discussion}

To demonstrate the role of the relativistic effects
in the amplitudes of the
$\chi_{_{0}}\rightarrow\psi\gamma$,
$\chi_{_{2}}\rightarrow\psi\gamma$ and
$\psi'\rightarrow\chi_{_{0}}\gamma$,
$\psi'\rightarrow\chi_{_{2}}\gamma$ transitions
we have presented in table 1 the ratios of the
overlap integrals for these transitions
extracted from experiment [12]
using the nonrelativistic formulas and
the formulas (32,33), (35,36).
It is seen that the relativistic corrections
change the ratios
$r_{\psi\chi_{_{0}}}/r_{\psi\chi_{_{2}}}$ and
$r_{\chi_{_{0}}\psi'}/r_{\chi_{_{2}}\psi'}$
in opposite directions. This
seems to be right from the point of view of
agreement
between the quark model predictions [4,6]
and experiment.
However, the experimental errors are large
and for strict conclusions new
measurements are necessary.

The results for the
$\psi\rightarrow\eta_{_{c}}\gamma$ transition
are presented in table 2.
The typical for the potential
models value: $m=1.6~GeV$ [4-6,13-15],
is used for the c-quark mass.
To demonstrate the dependence on this mass
the results obtained with
$m=1.8~GeV$ are also presented.
For $\alpha^{2}/m^{2}$ the value
giving the correct mean magnitude of $q^{2}/m^{2}$ in $\psi$:
$\alpha^{2}/m^{2}=\frac{4}{3}\langle q^{2}/m^{2}\rangle_
{\psi}\cong 0.3$ [13-15] is taken.
It is seen that the relativistic corrections
act in right direction leading to a suppression
of $\Gamma(\psi \rightarrow \eta_{_{c}}\gamma)$;
however, they are insufficient to remove
the existing disagreement of the nonrelativistic
quark model prediction with
experiment. The disagreement can be removed, if we take
larger c-quark masses, $m>1.8~GeV$, which can be accommodated
by introducing a negative additive constant in the potential.
However, to make
a reliable conclusions on the c-quark mass, we need
improved measurements of $\Gamma(\psi\rightarrow\eta_{_{c}}\gamma)$
with a better statistics.
Such measurements are an important experimental goal,
as the QCD sum rules predictions [16,17] (see table 2) also
exceed the experimental data.

$\bf{Acknowledgments}$.
I would like to thank R.A.Babayan, who took part
in the initial stage of this investigation,
and to V.O.Galkin and N.L.Ter-Isaakyan
for helpful discussions.
This work is supported by
Armenian Foundation of Scientific Researches (Grant
$\#~94-681$).\\

$\bf{References}$\\
1.  T.A.Armstrong et al., Phys.Rev.Lett. 68 (1992) 1468.\\
2.  T.A.Armstrong et al., Phys.Rev. D47 (1993) 772.\\
3.  T.A.Armstrong et al., Phys.Rev.Lett. 69 (1992) 2337.\\
4.  P.Mokhay, J.L.Rosner, Phys.Rev. D28 (1983) 1132.\\
5.  S.N.Gupta, S.F.Radford, W.W.Repko, Phys.Rev. D31 (1985) 160.\\
6.  S.Godfrey, N.Isgur, Phys.Rev. D32 (1985) 189.\\
7.  M.V.Terentiev, Yad.Fiz. 24 (1976) 207.\\
8.  V.B.Berestetski, M.V.Terentiev, Yad.Fiz. 24 (1976) 1044.\\
9.  I.G.Aznauryan, A.S.Bagdasaryan, N.L.Ter-Isaakyan,
Phys.Lett. 112B (1982) 393.\\
10. I.G.Aznauryan, A.S.Bagdasaryan, N.L.Ter-Isaakyan,
Yad.Fiz. 36 (1982) 1278.\\
11. I.G.Aznauryan, A.S.Bagdasaryan, Yad.Fiz. 41 (1985) 249.\\
12. "Review of Particle Properties," Phys.Rev. D50 (1994) 1173.\\
13. H.Crater, P.Van Alstine, Phys.Lett. 100B (1981) 166.\\
14. R.W.Childers, Phys.Lett. 126B (1983) 485.\\
15. J.L.Basdevant, S.Boukraa, Z.Physik C-Particles
and Fields 28 (1985) 413.\\
16. A.Yu.Khodjamirian, Yad.Fiz. 39 (1984) 970.\\
17. V.A.Beilin, A.V.Radyushkin, Nucl.Phys. B260 (1985) 61.

\newpage

{\large{\bf Table 1.~~~~~~}}
The ratios of the wave-function overlap integrals for the
$\chi_{0} \rightarrow \psi \gamma,~
\chi_{2} \rightarrow \psi \gamma$ and
$\psi' \rightarrow \chi_{0}\gamma,
\psi' \rightarrow \chi_{2}\gamma$
transitions
extracted from experiment[12]
in comparison with the quark model predictions[4,6]:
(a) the results extracted using nonrelativistic formulas,
(b) the results extracted taking
into account relativistic corrections
by the formulas (32,33) and (35,36).

\vspace {5mm}
\begin{center}
\begin{tabular}{|l|c|c|p{2.3cm}|p{2.3cm}|}
\cline{1-5}
Overlap   &  &&\multicolumn{2}{c|}{Quark model predictions} \\
integral  &(a)         &(b)        &    \multicolumn{2}{c|}{} \\
\cline{4-5}
ratios    &         &       & \qquad[4]     &\qquad[6]      \\
          \cline{1-5}
         &         &          &                 &            \\
\quad$\frac{r_{\psi \chi_{0}}}{r_{\psi \chi_{2}}}$ &
$0.98\pm 0.23$&$1.06 \pm 0.25$& \qquad $1.03$&$ \qquad1.02$  \\
         &         &                 &                 &            \\
\cline{1-5}
         &         &                 &                 &            \\
$\quad\frac{r_{\chi_{0}\psi'}}{r_{\chi_{2}\psi'}}$ &$0.83 \pm 0.05$ &
$0.77 \pm 0.05$& \qquad $0.81$ & \qquad$0.76 $  \\
& & & &\\ \cline{1-5}
\end{tabular}
\end{center}

\vspace {7mm}
{\large{\bf Table 2.~~~~~~}}
Our results for $\Gamma(\psi \rightarrow \eta_{c}\gamma)$
in comparison with QCD sum rules predictions and experiment.
In parentheses the nonrelativistic quark model predictions
are given.

\vspace {5mm}

\begin{center}
\begin{tabular}{|l|c|c|c|}
\hline
\multicolumn{2}{|c|}{Our results} &QCD sum rules& Experiment \\
\hline
$m=1.6 GeV$   &$ m=1.8 GeV$        &$1.8-2.0 KeV$           &      \\
\cline{1-2}
$2.06 KeV$         & $1.63 KeV$        & [16]     & $1.14\pm 0.3 KeV$\\
$(2.63 KeV)$ &$(2.06 KeV)$&$2.6 \pm 0.5 KeV$&$[12]$  \\
  &  & [17] & \\
\hline
\end{tabular}
\end{center}

\end{document}